# En-route to the fission-fusion reaction mechanism: a status update on laser-driven heavy ion acceleration

F. H. Lindner[1,a)], E. McCary[2], X. Jiao[2], T. M. Ostermayr[1,3], R. Roycroft[2,4], G. Tiwari[2], B. M. Hegelich[2,5], J. Schreiber[1], P. G. Thirolf[1]

[1]Lehrstuhl für Experimentalphysik – Medizinische Physik, Fakultät für Physik, Ludwig-Maximilians-Universität München, Am Coulombwall 1, 85748 Garching b. München, Germany

[2]Center for High Energy Density Science, University of Texas, Austin, TX 78712, USA

[3]Max-Planck-Institut für Quantenoptik Garching, Hans-Kopfermann-Str. 1, 85748 Garching b. München, Germany

[4]Los Alamos National Laboratory, Los Alamos, New Mexico 87545, USA

[5]Center for Relativistic Laser Science, Institute for Basic Science (IBS), Gwangju 61005, South Korea

**The fission-fusion reaction mechanism was proposed in order to generate extremely neutron-rich nuclei close to the waiting point N = 126 of the rapid neutron capture nucleosynthesis process (r-process). The production of such isotopes and the measurement of their nuclear properties would fundamentally help to increase the understanding of the nucleosynthesis of the heaviest elements in the universe. Major prerequisite for the realization of this new reaction scheme is the development of laser-based acceleration of ultra-dense heavy ion bunches in the mass range of A = 200 and above. In this paper, we review the status of laser-driven heavy ion acceleration in the light of the fission-fusion reaction mechanism. We present results from our latest experiment on heavy ion acceleration, including a new milestone with laser-accelerated heavy ion energies exceeding 5 MeV/u.**

---

a)florian.lindner@physik.lmu.de

## 1. INTRODUCTION

The nucleosynthesis of the heaviest elements in the universe follows the rapid neutron capture process (r-process) at astrophysical sites like binary neutron star mergers [1]. The r-process passes through the neutron-rich side of the chart of nuclides. The generation of the involved isotopes, especially around the waiting point at the magic neutron number N = 126, is still out of reach for conventional accelerators. Hence, the nuclear properties of a large fraction of the r-process nuclei are presently unknown and cannot be studied in the laboratory.

Habs et al. proposed to exploit laser-driven ion acceleration for the production of neutron-rich nuclei close to this waiting point and introduced the fission-fusion reaction mechanism [2]. This is a two-step process, based on the fission of heavy ions like thorium and the subsequent fusion of the neutron-rich fission fragments. Due to the limited density of conventionally accelerated ion bunches, this reaction mechanism is hardly accessible using existing conventional accelerator facilities, where the fusion probability would be negligible. In contrast, short pulse laser-based acceleration of heavy ions can approach solid-state-like density of the ion bunches, when realizing the acceleration in the Radiation Pressure Acceleration (RPA) regime [3–6]. Owing to the unprecedented high density of the laser-generated heavy ion bunches, the yield of the fission-fusion process is expected to reach usable reaction product numbers.

Thus, the development of laser-driven heavy ion acceleration in the RPA regime is a major prerequisite for the realization of the fission-fusion reaction mechanism. A second constraint is placed by the minimum energy required to overcome the fission barrier: the heavy ions need to achieve kinetic energies above ca. 7 MeV/u. Whilst laser-driven acceleration of protons and light ions has been extensively studied for about two decades [7–9], the acceleration of ions with mass numbers around A $\approx$ 200 and above is still sparsely investigated.

To our knowledge, only two papers contain experimental data on laser-driven heavy ion acceleration in the mass and energy range relevant to the fission-fusion reaction mechanism. The first one was published in 2000 by Clark et al. and reports on lead ions, laser-accelerated up to energies around 2 MeV/u [10]. The ions originated from the front side of 2 mm thick lead targets, which were irradiated with a ps glass laser system with a pulse energy of 50 J and an intensity of $5 \times 10^{19}$ W/cm². In the second relevant publication from 2015, Braenzel et al. focused a Ti:sapphire laser pulse with an intensity of $8 \times 10^{19}$ W/cm² and an excellent laser contrast onto 14 nm thick gold foils and thereby accelerated gold ions to maximum energies of 1 MeV/u [11].

Besides these experimental publications, two simulation papers were published by Petrov et al. in 2016 and 2017, containing promising theoretical studies on gold ion acceleration. They investigated different acceleration mechanisms and their influence on the ion beam parameters by varying the gold foil thickness [12] and the laser pulse duration [13]. In particular, they observed indications for RPA with short-pulse laser systems, delivering gold ions bunches with energies far beyond the fission barrier of 7 MeV/u. Hence, these

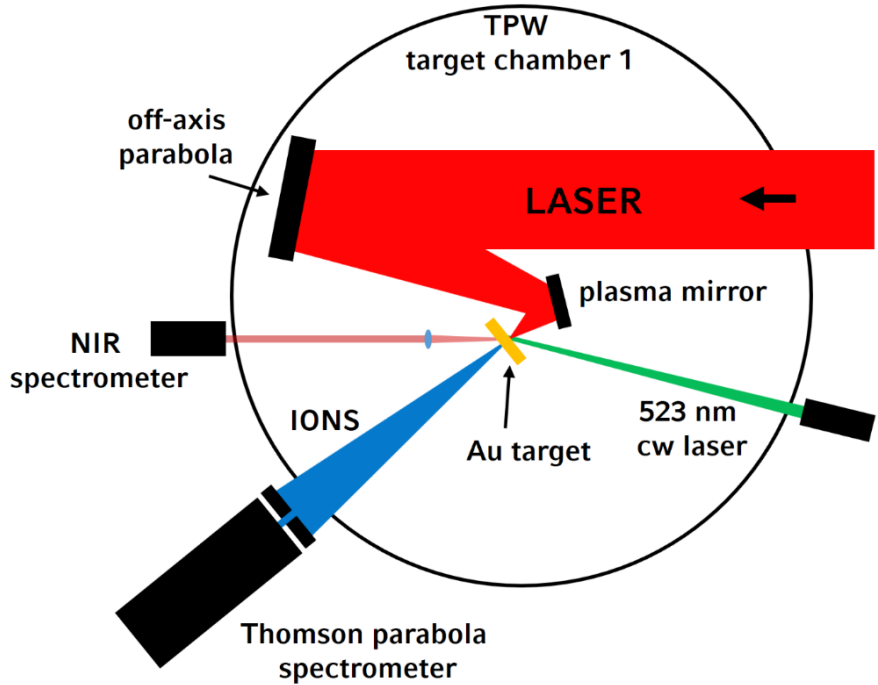
*Figure 1. Experimental setup at the Texas Petawatt laser (TPW).*

2D PIC simulation results make a strong case for soon fulfilling the demands on laser-driven heavy ion acceleration for the fission-fusion reaction mechanism [2,14].

However, especially for the development of RPA, which would ideally deliver a high-density, narrow-bandwidth heavy ion bunch around 7 MeV/u, we will have to rely on the upcoming new generation of high-power laser systems, like the 2 x 10 PW laser which is currently under construction at the Extreme Light Infrastructure – Nuclear Physics (ELI-NP) facility in Măgurele near Bucharest [15]. In the meantime, it is inevitable to gain knowledge on and control over heavy ion acceleration with the already available laser systems. In this paper, we report about our latest experimental progress on laser-based acceleration of heavy ions. We present the energy spectra of laser-accelerated gold ions originating from thin foils with varying thickness. Compared to the early studies of laser ion acceleration [10], we observe heavy ion energies that are a factor of 2.5 higher than reported before. Furthermore, we show the influence of radiative target heating on the ion energies and numbers.

## 2. EXPERIMENTAL SETUP

The experiment was performed at the Texas Petawatt laser (TPW) of the University of Texas at Austin [16], which delivered about 110 J laser pulse energy within 140 fs at a central wavelength of 1057 nm. The setup is sketched in Figure 1. The TPW was focused by an off-axis parabola to a focal spot size of 10 µm. A single inline plasma mirror was used for contrast enhancement, which resulted in a laser intensity of $(8 \pm 2) \times 10^{20}$ W/cm² on target.

We employed freestanding gold foils with thicknesses of 50 nm, 100 nm and 300 nm as targets. The foils were manufactured by the LMU target factory by physical vapor deposition. Hydrocarbons exist even in clean vacuum environments and will naturally accumulate on the target surface as contaminants. Due to their higher charge-to-mass ratio compared to heavier ions, they can strongly suppress or cancel heavy ion acceleration [17]. Therefore, we used a continuous wave laser with a wavelength of 532 nm and an adjustable output power up to 1 W for radiative target heating in order to remove the surface contaminants. The foils were heated above 500 °C, while the temperature was monitored by a commercial NIR spectrometer [18], which measured the spectrum of the thermal radiation of the laser-heated gold target foils.

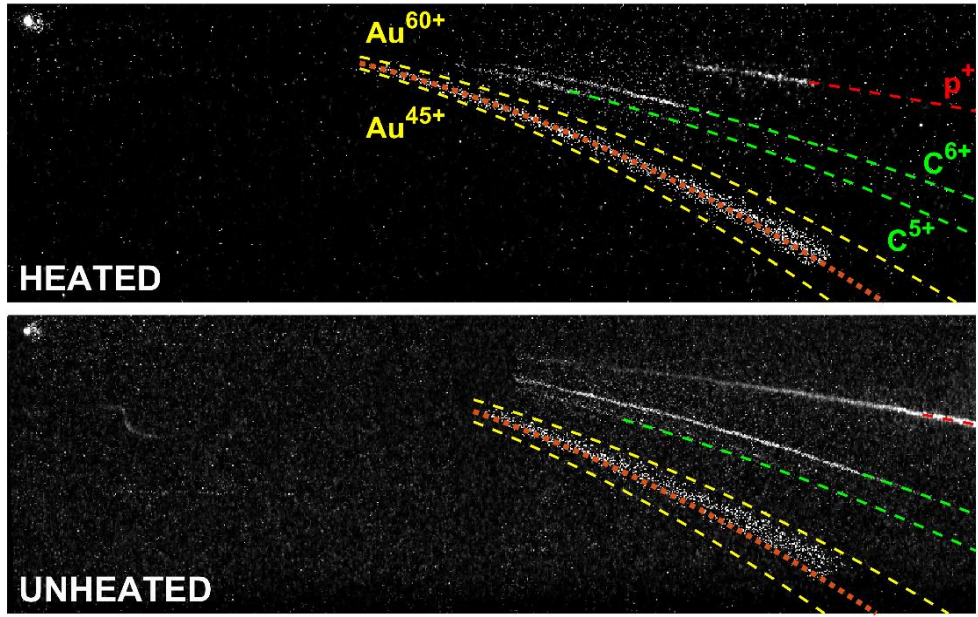

*Figure 2. Raw data from the Thomson parabola spectrometer from heated (top) and unheated (bottom) shots on 100 nm thick gold foils. The gold pits (between the yellow parabolas, depicting the parabolas for $Au^{45+}$ and $Au^{60+}$) and the carbon curves (green) are extracted from CR39 track detectors, while the proton curves (red) were acquired using image plates, which were located behind the CR39 plates. Different scaling factors have been applied to the individual lines in order to enable a reasonable comparison: the proton (carbon) data for the heated shot has been multiplied with a factor of 6.4 (factor of 2) compared to the unheated shot. The orange line indicates the charge state $Au^{51+}$.*

As ion diagnostics, we employed a Thomson parabola spectrometer [19] in the target normal direction with a 200 µm entrance pinhole, accepting a solid angle of $2 \times 10^{-5}$ msr of the emitted beam. CR39 nuclear track detector sheets with a thickness of 1 mm in front of image plates (IPs) served as passive ion detectors. The CR39 was chosen as heavy ion detector, as a distinction between heavy particles (gold ions) and light particles (carbon ions) is easily possible by the different pit sizes.

3. RESULTS AND DISCUSSION

Figure 2 shows exemplarily raw data taken with the Thomson parabola spectrometer from a heated (top) and an unheated (bottom) shot on 100 nm thick gold foils. The red and green lines indicate the analytically calculated parabolas for protons as well as for $C^{6+}$ and $C^{5+}$, respectively. Single gold traces are not resolved. The observed impacts of gold ions are enclosed by the calculated parabolas for $Au^{45+}$ and $Au^{60+}$ (yellow lines). The gold and carbon impacts have been recorded with CR39, while the proton trace was taken from IPs, positioned directly behind the CR39. The figure is an overlay of three images: manually counted gold pits, automatically registered carbon pits and the IP image with the proton trace. The orange line indicates the parabola for $Au^{51+}$, which is according to references [12,13] expected to be the most prominent charge state for the employed laser intensity, which is in rough agreement with our measurements for thin gold targets.

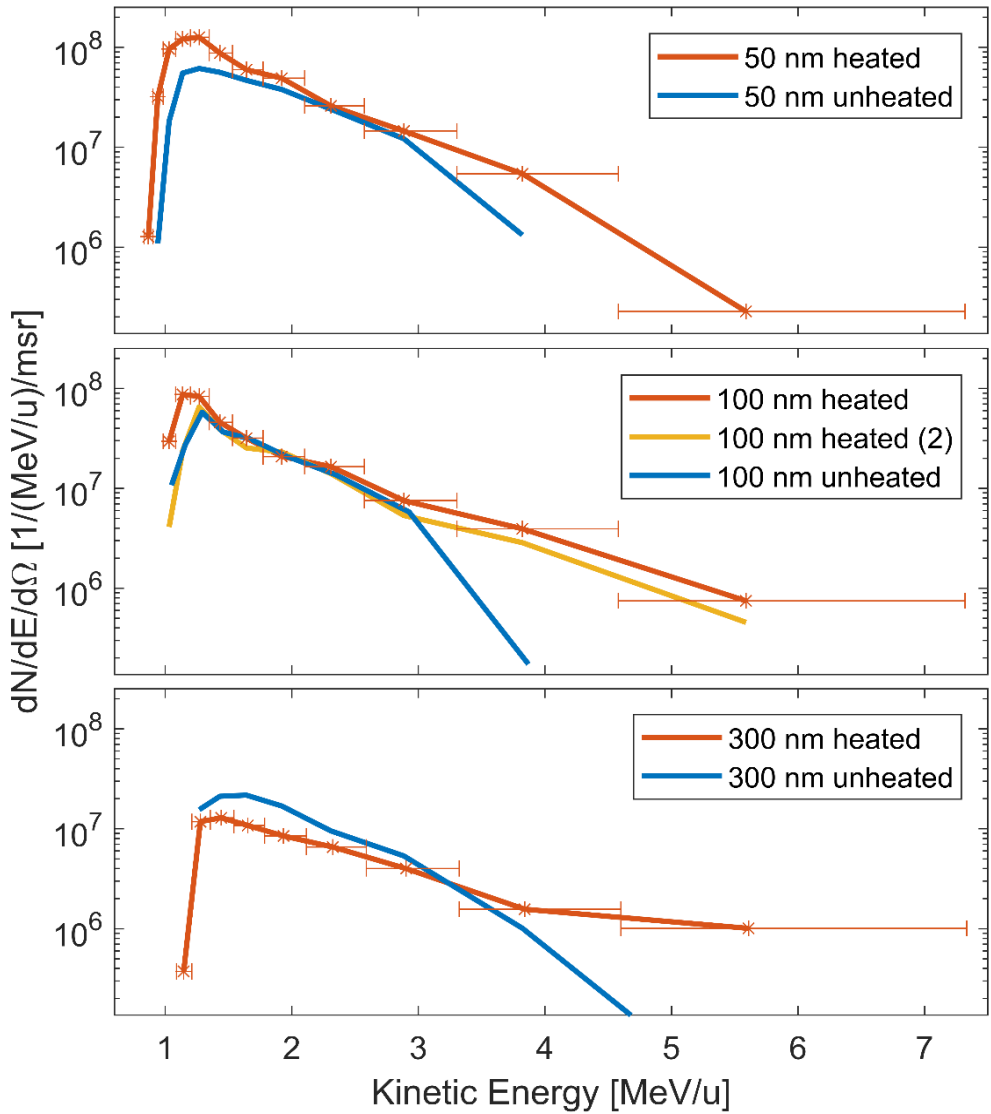

*Figure 3. Gold ion spectra from shots on heated (red, yellow) and unheated (blue) gold foils with varying thickness (top: 50 nm, middle: 100 nm, bottom: 300 nm). The energy spectra are integrated over all charge states. The yellow line corresponds to the same temperature as the red lines.*

The still appearing protons in Figure 2 indicate that the target heating was insufficient to remove all proton contaminants. Follow-up studies for a more efficient target cleaning are already in progress. Meanwhile, the proton and carbon ion numbers are significantly reduced in heated shots. The beneficial effect on gold ion energies is evident, which indicates that target heating in general is an effective method.

Figure 3 shows the gold ion spectra for target foil thicknesses of 50 nm, 100 nm and 300 nm. The spectra are integrated over all charge states, as the Thomson parabola spectrometer did not resolve single gold ion parabolas. For each thickness, spectra from shots on heated (red, yellow) and on unheated (blue) targets are presented. Clearly, the energy spectra are monotonically decaying, indicating the predominant acceleration in the Target Normal Sheath Acceleration (TNSA) regime [20]. While the shape of the spectra does not depend on the target thickness, the number of gold ions, which have been measured in target normal direction, increases with decreasing gold foil thickness. A clear difference between the energy spectra from heated and unheated shots is observable: the spectra from unheated shots decrease much steeper towards higher energies. The highest energies are achieved for shots on heated targets and exceed 5 MeV/u, independent of the target thickness. This measured maximum energy exceeds the highest kinetic energies, which have been reported so far for laser-accelerated heavy ions in this mass range [10], by a factor of about 2.5. The drop at energies below 1 MeV/u is caused by the geometry of the spectrometer.

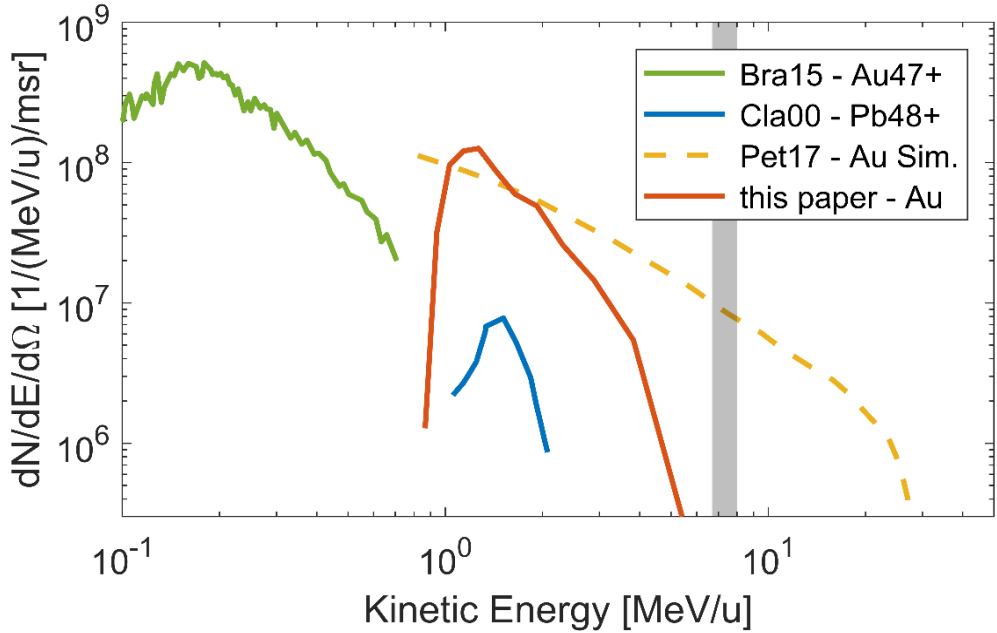

*Figure 4. Comparison of different heavy ion spectra. The green curve is the gold data taken by Braenzel et al. in 2015 [11]; blue curve corresponds to lead ions, measured by Clark et al. in 2000 [10]. The red curve is our measurement at the TPW laser for a heated, 50 nm thick gold foil in comparison to a spectrum simulated for the TPW parameters and a gold foil thickness of 20 nm by Petrov et al. [13]. The grey shaded area indicates the rough energy range of the fission barrier for $^{232}$Th around 7 MeV/u, which to overcome is a basic prerequisite for the fission-fusion reaction mechanism.*

A direct comparison of our measurement with laser-accelerated heavy ion spectra in literature (see introductory section) is shown in Figure 4. The green and blue curves show previous measurements by Braenzel et al. [11] and Clark et al. [10], respectively. The red curve shows our measurement for a heated, 50 nm thick gold foil, which represents an important step towards the indicated fission-fusion goal of about 7 MeV/u for $^{232}$Th (grey shaded area). The yellow, dashed line shows a gold ion spectrum simulated by Petrov et al. [13]. For this simulation, they used a gold target with a thickness of 20 nm covered with a 5 nm thick $H_2O$ contamination layer on the rear side and laser parameters similar to the TPW laser (50 J, 180 fs, 5µm FWHM focal spot size, $1 \times 10^{21}$ W/cm² laser intensity).

Comparing the gold spectra from the TPW laser (measured) and simulation results for similar conditions reveals that the simulation overestimates both ion energy and particle numbers.

Our measurements show an increasing number of ions in the target normal direction towards thinner targets, which is likely to show an increased directionality of the ion bunch with decreasing foil thickness. However, 20 nm thick targets were not available in the presented experiment and the ion bunch profile was not recorded. Thus, a direct comparison of the ion numbers cannot be provided. Concerning the gold ion energies, our measurements imply a behavior independent of the target thickness. It is most likely, that the energies required for the fission-fusion reaction mechanism will be achievable at more powerful laser systems.

Further studies on laser-driven heavy ion acceleration are already planned using high-power lasers with different pulse characteristics, like the PHELIX laser at the GSI in Darmstadt (200 J, 500 fs) and the ATLAS 3000 at the Centre for Advanced Laser Applications (CALA) in Garching near Munich (60 J, 20 fs) [14] with a peak intensity around $10^{22}$ W/cm². These experiments are preparations for studies at the new high-power laser at ELI-NP (~$10^{23}$ W/cm²) [15]. In particular, the transition from TNSA to RPA could come into reach for these laser intensities, paving the way to a more efficient acceleration of large ion numbers to higher energies, as they are required for the fission-fusion reaction mechanism.

## 4. CONCLUSION

We presented a status update on laser-driven heavy ion acceleration in the light of the fission-fusion reaction mechanism, which aims at generating neutron-rich heavy ions close to the r-process waiting point around N = 126. We showed promising results from a recent gold ion acceleration campaign at the Texas Petawatt laser. In particular, we measured experimentally gold ions with energies of more than 5 MeV/u. The results and experiences support our optimism that the energetic requirements of the ion bunches for the fission-fusion reaction mechanism can be met with the soon operational next-generation laser systems.

## ACKNOWLEDGEMENTS

The authors acknowledge funding by the BMBF under contract 05P15WMEN9 and contract 05P18WMEN9, the DFG Cluster of Excellence MAP (Munich-Centre for Advanced Photonics) and the Centre for Advanced Laser Applications. The authors thank the TPW facility staff for operating the laser and their strong support during the whole campaign. This work has been carried out within the framework of the EUROfusion Consortium and has received funding, through the ToIFE, from the European Union's Horizon 2020 research and innovation program under grant agreement number 633053.